\documentclass[%
 reprint,
 twocolumn,
 amsmath,amssymb,
 aps,
 prl,
 nofootinbib,
 tightenlines,
 nobibnotes
]{revtex4-2}

\usepackage{graphicx}
\usepackage{dcolumn}
\usepackage{bm}
\usepackage{hyperref}

\usepackage{color}

\usepackage{gensymb}

\definecolor{Green}{RGB}{0,100,0}
\definecolor{Purple}{RGB}{102,0,255}
\definecolor{Blue}{RGB}{51,153,255}
\definecolor{Red}{RGB}{151,010,010}

\definecolor{Orange}{RGB}{255,69,0}

\begin{document}

\title{Optical Vortex Braiding with Bessel Beams}

\author{Andrew A. Voitiv$^1$, Jasmine M. Andersen$^1$, Mark E. Siemens$^1$}
\email[]{msiemens@du.edu}%
\author{Mark T. Lusk$^2$}%
\email[]{mlusk@du.edu}%
\affiliation{$^1$Department of Physics and Astronomy, University of Denver, 2112 E. Wesley Avenue, Denver, CO 80208, USA
}%
\affiliation{
$^2$Department of Physics, Colorado School of Mines, 1500 Illinois Street, Golden, CO 80401, USA
}%

\date{\today}

\begin{abstract}
We propose the braiding of optical vortices in a laser beam with more than $2\pi$ rotation by superposing Bessel modes with a plane wave. We experimentally demonstrate this by using a Bessel-Gaussian beam and a coaxial Gaussian, and we present measurement of three complete braids. The amount of braiding is fundamentally limited only by the numerical aperture of the system and we discuss how braiding can be controlled experimentally for any number of vortices.
\end{abstract}

\maketitle

Optical vortices are phase singularities in light fields that are characterized by a phase wrap of $2\pi \ell$, with topological charge $\ell$, which trace out lines of darkness as the field propagates. They can be found in laser speckle and random waves \cite{Berry2000PhaseWaves} and in eigenmodes of the Helmholtz equation in free space for laser beams, such as Laguerre-Gaussian and Bessel modes \cite{Allen1992OrbitalModes,Durnin1987ExactTheory}. Laser beams containing a single, centered vortex with integer-valued orbital angular momentum (OAM) have been extensively studied and engineered, but beams with two or more optical vortices have received less attention. Nevertheless, the dynamics of optical systems with multiple vortices is a rich field of study, which has given rise to overlaps and analogues with other physical systems \cite{Alperin2019QuantumLight,Shen2019OpticalSingularities} inspired by the hydrodynamical nature of vortices. For few-optical-vortex fields, authors have studied various linked, knotted, and braided vortex loops \cite{Berry2001KnottedSingularities,Dennis2003BraidedSuperpositions,Dennis2010IsolatedKnots}, including experimental realizations for knots and links \cite{Dennis2010IsolatedKnots,Zhong2019AccurateKnots}. In addition to contributing to the study of optical vortex dynamics in structured laser light, vortex braiding could find potential application, for example, in microstructure fabrication \cite{Ni2017Three-dimensionalMaterial}. In this Letter, we propose and experimentally demonstrate a scheme to controllably braid optical vortices in a laser beam in free space, with composite Bessel-Gaussian modes. We experimentally measure three full vortex braids in a propagation distance of $61$ cm.

\begin{figure}[ht!]
\centering
\includegraphics[width=.83\linewidth]{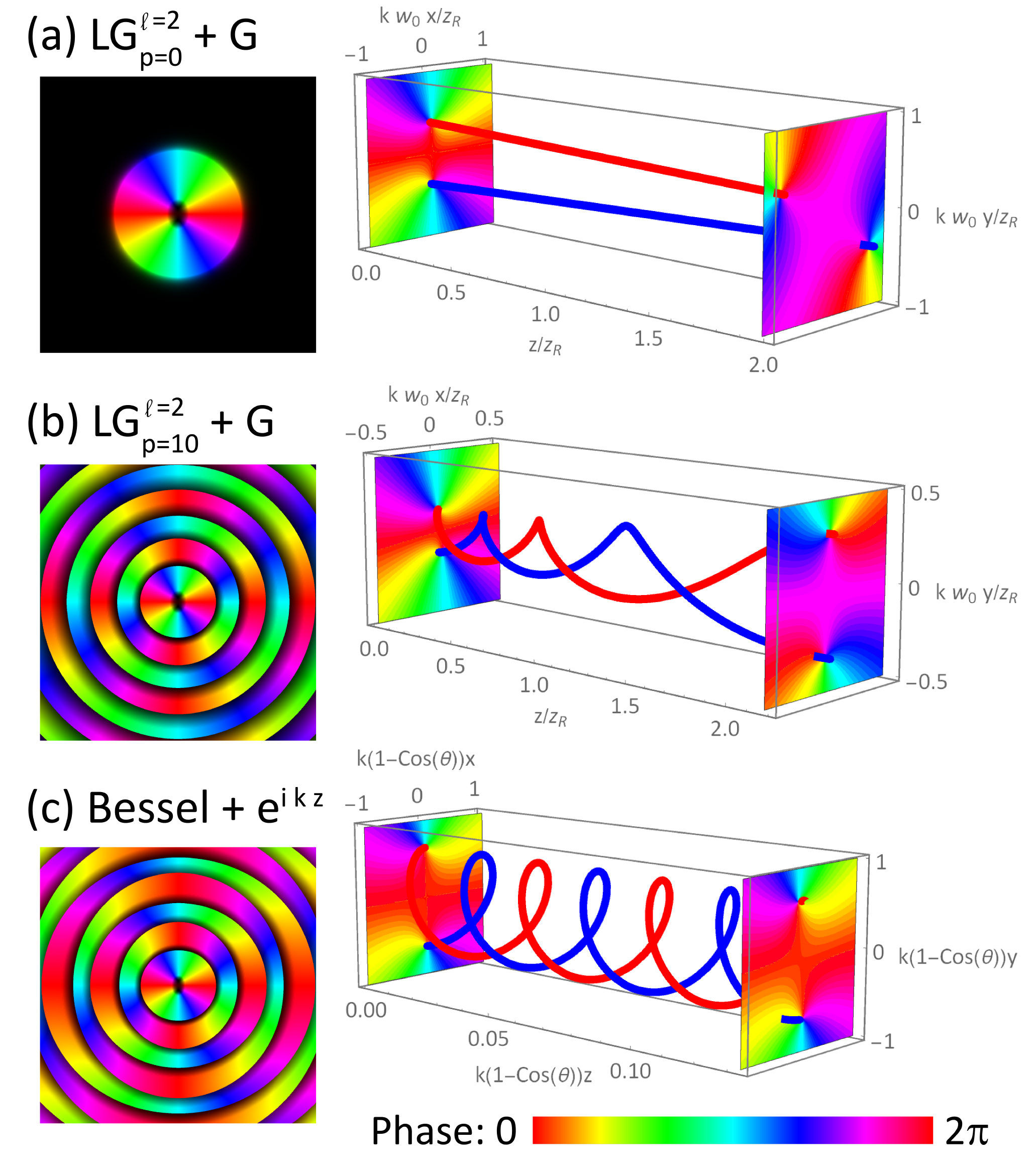}
\caption{(Left) Combined phase (hue) and amplitude (brightness) transverse profiles of two-vortex laser modes and (Right) the associated vortex trajectories. (a) A Laguerre-Gaussian of $\ell=2$ and $p=0$ superposed with a collinear Gaussian results in two optical vortices that move in straight lines to a maximum quarter rotation. (b) By increasing the radial number, here $p=10$, the vortices rotate and braid over large propagation distances. (c) A Bessel of topological charge $\ell=2$ superposed with a plane wave facilitates vortex braiding. The divergence of LG beams is evident in the increased separation distance between the two vortices, whereas the separation is constant in (c) due to the divergenceless property of ideal Bessel beams. A Gaussian weight of 0.05 is used in all cases.}
\label{fig:limitation}
\end{figure}

Optical vortex braiding refers to the rotation of two or more vortices around the central axis of propagation in an optical beam; the rotation is periodic, and hence at least one complete $2\pi$ rotation should be obtained. A natural first step to attempt braiding is with Laguerre-Gaussian (LG) vortex beams, as these are widely familiar and accessible. It is well known that an LG beam with a central optical vortex of charge $|\ell| \geq 1$ can be split into $|\ell|$ vortices, each of charge magnitude $1$, by superposing a Gaussian or plane wave. For example, adding a Gaussian to a coaxial LG mode with $\ell=2$ yields two $\ell=1$ vortices shifted symmetrically from the center of the beam. These vortices appear to rotate with propagation, as shown in Figure \ref{fig:limitation}(a), due to the difference in Gouy phase between the modes. The Gouy phase is expressed as: $\psi(z) = (|\ell| + 2p + 1) \arctan \left( z / z_R \right)$, for propagation distance $z$, Rayleigh range $z_R$, and radial mode number $p$. As in the case of optical vortices within a Gaussian background field \cite{Rozas1997PropagationVortices}, this Gouy phase limits the rotation in Figure \ref{fig:limitation}(a) to $\pi / 2$.

Our work is motivated by the observed emergence of braiding, with increasing radial mode number $p$, in composite LG-Gaussian beams \cite{siegman1986lasers}. For instance, the case of $p=10$ is illustrated in Figure \ref{fig:limitation}(b), which braids twice (two complete revolutions). However, the braiding period associated with this composite is unrealistically large; a beam waist of $w_0 = 3.8$ mm, wavelength $\lambda = 633$ nm, and Rayleigh range $z_R = 72$ m requires a propagation distance of $z = 144$ m to achieve the two braids shown. Further improvement is obtained by first increasing the LG radial number to the limit of $p \rightarrow \infty$, which approximates as a Bessel-Gaussian beam. An ideal Bessel mode can be written as \cite{Durnin1987ExactTheory}:
\begin{equation} \label{bessel}
    E(\rho,\phi,z) = J_{\ell}(k_{\rho} \rho) e^{i \ell \phi} e^{i k_z z}.
\end{equation}
Here $z$ is the propagation coordinate, $(\rho,\phi)$ are the polar coordinates in the transverse plane, $J_{\ell}$ is an $\ell$-th order Bessel function of the first kind, the radial wave number of the Bessel is $k_{\rho} = k \sin{\theta}$, and the longitudinal wave number is $k_z = k \cos{\theta}$. A Bessel beam is equivalent to a sum of plane waves propagating on the surface of a cone \cite{Mcgloin2005BesselLight}, where $\theta$ is the angle subtended from the z-axis defining this cone: $\theta = \arcsin{\left( k_{\rho} / k \right)}$. 

Figure \ref{fig:limitation}(c) demonstrates much faster vortex braiding by superposing a copropagating plane wave, $e^{i k z}$, with a Bessel beam of topological charge $\ell=2$. As detailed next, closed-form expressions can be derived for the vortex position and braiding period.

Beginning with the paraxial approximation \cite{Lax1975FromOptics}, we take Equation \ref{bessel} as our particular form of the paraxial field and add a scattering plane wave to separate the $\ell$-th order Bessel vortex into $\ell$ first order vortices:
\begin{equation} \label{superposition}
    E(\rho, \phi, z, t) = \gamma e^{i(kz - \omega t)} + J_{\ell}(k_{\rho} \rho) e^{i (k \: \cos{\theta} \: z - \omega t + \ell \phi)}.
\end{equation}
The plane wave scaling amplitude, $\gamma := J_{\ell}(k_{\rho} \rho)|_{\rho=\rho_0}$, shifts the vortices by $\rho_0$ from the beam center. The vortices are located at positions for which $E(\rho, \phi, z, t) = 0$, which are the cylindrical positions $(\rho_0, \, \phi_{0,n})$, where each vortex is labelled by $n \in [1,\ell]$. The polar position of vortex $n$ is found to be:
\begin{equation} \label{polar}
    \phi_{0,n}(z) =  \frac{k}{\ell} (1 - \cos{\theta}) z  + (2n - 1) \frac{\pi}{\ell}.
\end{equation}
This analytical expression for the trajectories of all $\ell$ rotating vortices in the composite Bessel beam was used to create the braiding trajectory in Figure \ref{fig:limitation}(c). The braiding period, obtained directly from Equation \ref{polar}, is given by:
\begin{equation} \label{zbraid}
    z_{braid} = \frac{2 \pi \ell}{k (1 - \cos \theta)}.
\end{equation}

\begin{figure}[h!]
\centering
\includegraphics[width=\linewidth]{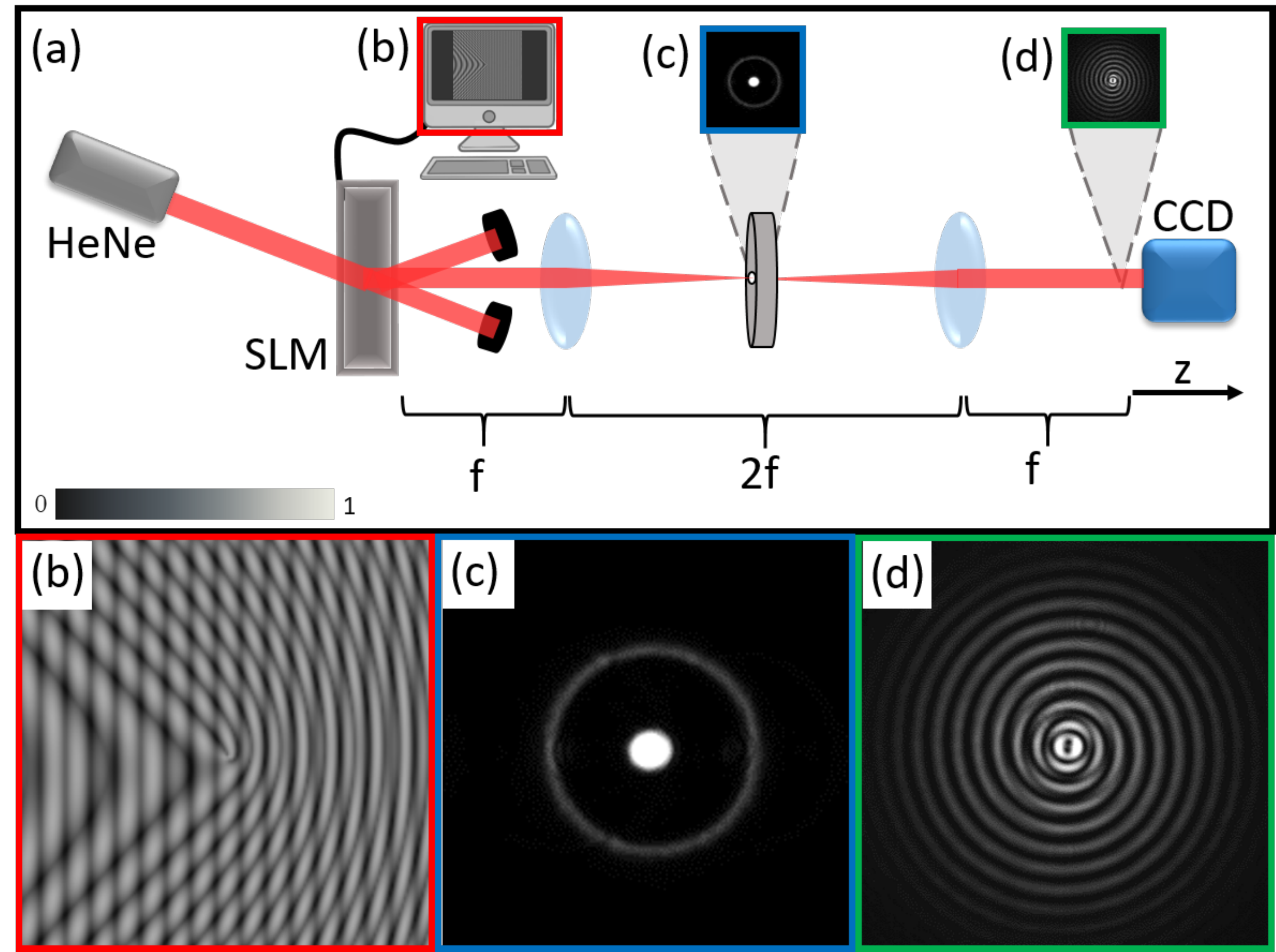}
\caption{(a) A HeNe laser passes through a hologram on an SLM, from which the first diffracted order is imaged and recorded by a CCD. (b) SLM display of a digital helical axicon grating used to generate a composite Bessel and Gaussian beam. (c) Image at the Fourier plane illustrates the finite lifetime of the Bessel-Gaussian. The central spot is the coaxial Gaussian. A pinhole blocks other diffracted orders not depicted. (d) Representative measurement of a Bessel-Gaussian beam between the imaging plane and $z_{max}$.}
\label{fig:schematic}
\end{figure}

Although Bessel beams are not physical, they can be well-approximated by Bessel-Gaussian beams  produced using an axicon \cite{Durnin1987ExactTheory, McLeod:54}. Their laboratory implementation has a finite aperture and is thus Bessel-like for only a finite distance, $z_{max} = w_0 / \tan{\theta}$, for beam waist $w_0$. Beyond this distance, the beam evolves into an annular ring as shown in the Fourier plane of Figure \ref{fig:schematic}(c).

The grating of a superposition of axicon and plane wave has the form:
\begin{equation} \label{hologram}
    \textrm{Hologram}(\rho,\phi) = \textrm{Abs}\left[e^{i \ell \phi + i k_{\rho} \rho} + C + e^{i k_g \rho \cos{\phi}} \right].
\end{equation}
Here the first term is the helical axicon, $C$ is the relative amplitude of the superposed plane wave (which yields a Gaussian experimentally), and the last term represents the tilted plane wave that acts as the reference for the hologram and produces a grating with spacing set by $k_g$. The topological charge of the Bessel-Gaussian mode is given by $\ell$ and its radial wave number $k_{\rho}$ is the slope of the axicon. The central vortex of the Bessel-Gaussian of charge $\ell$ is split into separate vortices by the superposed plane wave.

The experimental setup for producing and propagating composite Bessel-Gaussian modes is shown in Figure \ref{fig:schematic}(a). A collimated Helium Neon (HeNe) laser beam is incident on a spatial light modulator (SLM) which displays a grating as represented in Equation \ref{hologram}. Using two lenses and a pinhole, we can simultaneously image the mode of interest from the point of generation and also isolate the mode from the other diffracted orders off the grating. Starting at the imaging plane, one focal length from the second lens, a charged coupled device (CCD) sits on a motorized translation stage, allowing for measurement at increasing propagation distances. We use a beam ($\lambda = 633$ nm and $2$ mW of power) with a waist of $w_0 = 3.8$ mm and a transmissive SLM, which is a computer-controlled Epson 83H projector \cite{Huang2012ALabs}. The hologram of Equation \ref{hologram} is programmed with plane wave amplitude $C = 1$, radial wave number $k_{\rho} = 35,000$ $\textrm{rad} \cdot \textrm{m}^{-1}$, and topological charges $\ell = 2$ and $4$ for two different measurements.

\begin{figure}[h!]
\centering
\includegraphics[width=\linewidth]{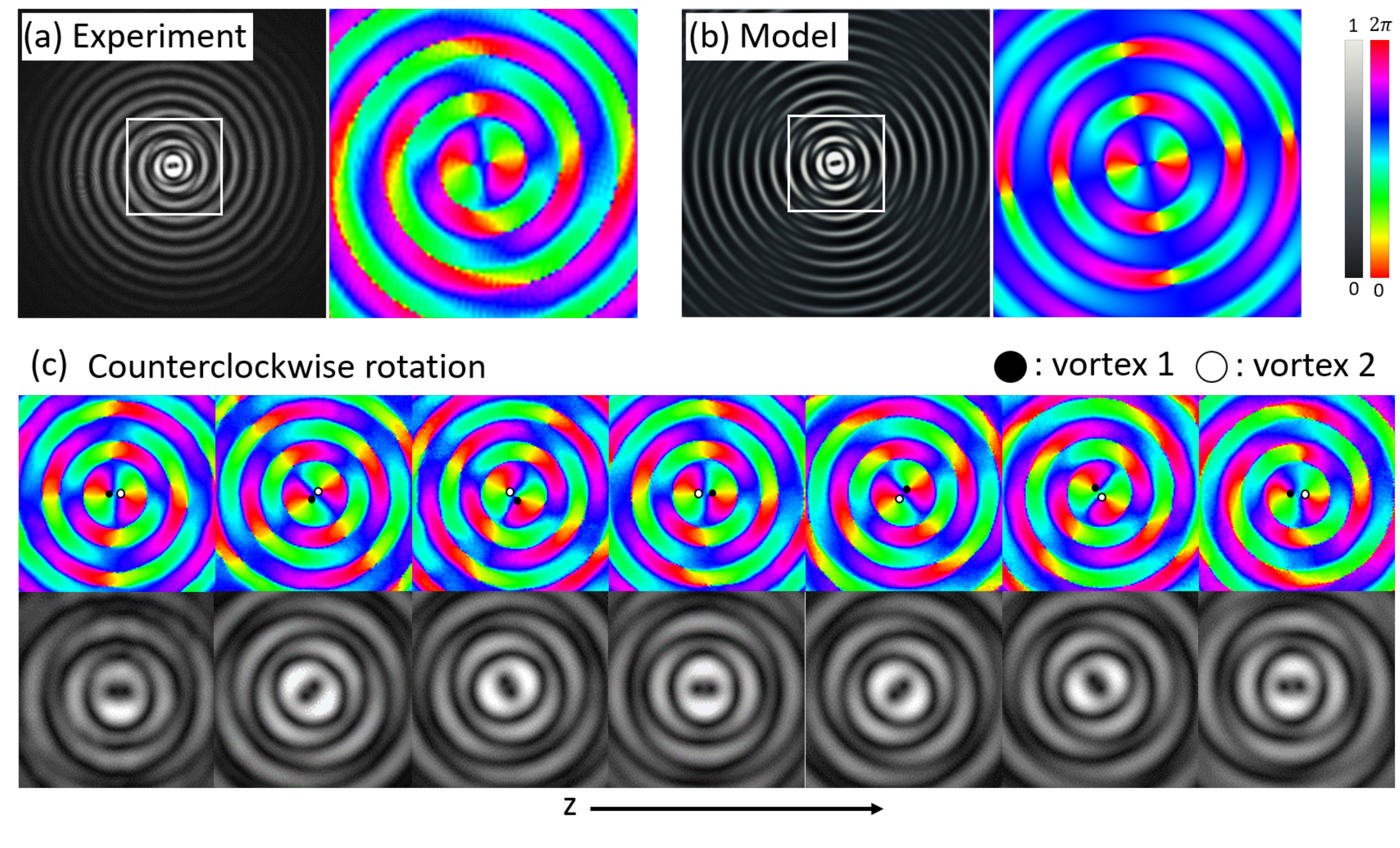}
\caption{(a) Experimentally acquired amplitude and phase of an $l=2$ Bessel-Gaussian beam superposed with a Gaussian. (b) Corresponding numerical prediction from Equation \ref{superposition}. The phase plots at right are plotted over the white-boxed region at left. (c) Measurements of phase and amplitude for one braiding period in $3.4$ cm steps from $z=15.2$ to $35.6$ cm.}
\label{fig:slices}
\end{figure}

Measurement of the transverse intensity at specified axial positions reveals the relative rotation of the optical vortices in the beam. While vortices are associated with dark spots in the optical field amplitude, they are intrinsically phase structures and are most naturally and directly identified in the phase of the beam. We chose to use phase-shifting digital holography in a collinear geometry to do this. Four interferometric images are acquired, each with a reference wave added to the hologram that has a phase shift of $\pi / 2$ between acquisitions \cite{Andersen2019CharacterizingHolography}. The four images are then combined algebraically to determine the phase \cite{Yamaguchi:97}. Figure \ref{fig:slices}(a) shows a representative resulting image and Figure \ref{fig:slices}(b) shows the corresponding numerical model of Equation \ref{superposition} ($t=0$) with the experimental parameters at associated propagation distance $z=46.5$ cm. The match between the Bessel-Gaussian measurement with the idealized Bessel model validates the predictions of Equations \ref{polar} and \ref{zbraid}.

\begin{figure}[htbp]
\centering
\includegraphics[width=\linewidth]{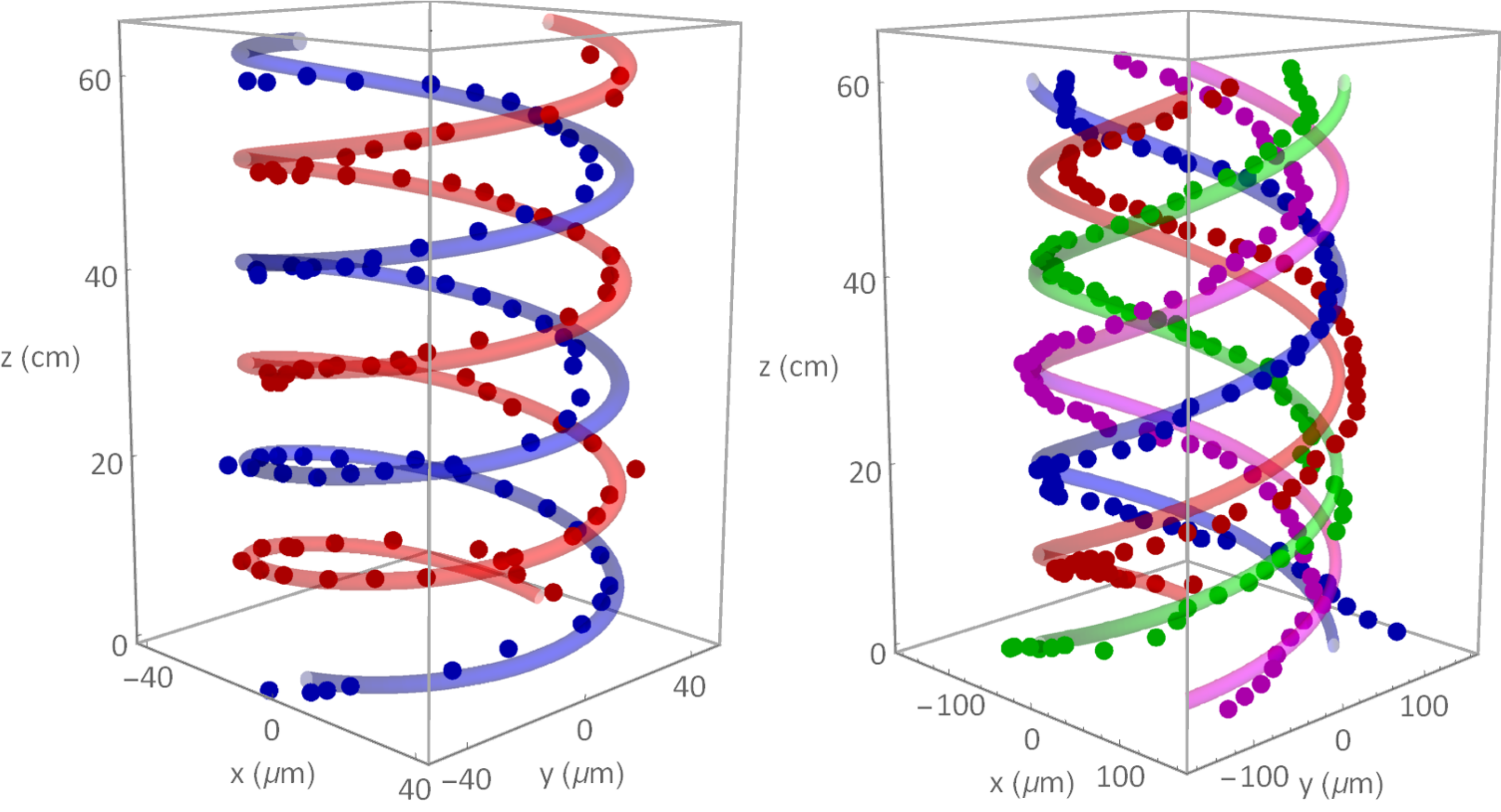}
\caption{Experimental and predicted braiding trajectories for (Left) $\ell = 2$ and (Right) $\ell = 4$. The points are experimentally-measured vortex locations and the continuous tubes \cite{Berger2001} are plotted from the model of Equation \ref{polar}. Vortices are distinguished by color. Radial distances, $\rho_0$, of $4.14 \times 10^{-3}$ cm and $1.21 \times 10^{-2}$ cm were used respectively, based on measurement from data and the model of Figure \ref{fig:slices}.}
\label{fig:noodles}
\end{figure}

The amplitude and phase data at each propagation step is subsequently used \cite{Dennis2009ChapterSingularities} to produce plots of vortex braiding as shown in Figure \ref{fig:noodles}, where the data was taken over a propagation distance of $61$ cm in $1$ cm steps. Equation \ref{zbraid} predicts that, for two vortices, there should be one complete braid every $20.4$ cm; for four vortices, $z_{braid} = 40.8$ cm.  For the two vortex case, the data qualitatively reveals three complete windings, consistent with the expression given by Equation \ref{zbraid}: $\frac{61 \: \textrm{cm}}{20.4 \: \textrm{cm}} \approx 3 \: (2.99)$. For the four vortex case, there are one and a half braids, again consistent with predictions: $\frac{61 \: \textrm{cm}}{40.8 \: \textrm{cm}} \approx 1.5 \: (1.495)$. Three braids is by no means a limit on the number of possible braids, and the methodology can be used to controllably braid any number of vortices.

The maximum number of braids for a composite Bessel-Gaussian beam can be determined by finding the angle $\theta$ that maximizes the ratio of $z_{max}$ to $z_{braid}$: $\frac{z_{max}}{z_{braid}} \propto \frac{1-\cos{\theta}}{\tan{\theta}}$, where the beam waist, wave number, and topological charge are held constant. The number of braids is thus maximized for an angle of $\theta = 0.9$ radians, which is controlled by the value of the radial wave number $k_{\rho}$. In the current arrangement, the number of braids could have been maximized by using $k_{\rho} = 7.8 \times 10^{6}$ $\textrm{rad} \cdot \textrm{m}^{-1}$. A value of $k_{\rho} = 35,000$ $\textrm{rad} \cdot \textrm{m}^{-1}$ was chosen instead because values higher than this produced gratings of Equation \ref{hologram} with variation in the phase which could not be represented by the finite pixels of the SLM, resulting in incomplete information imparted onto the incident beam. Furthermore, as the opening angle $\theta$ increases, the focusing effect of the axicon increases, which squeezes the transverse region of interest into a tighter space and thereby decreases the vortex separation. A very tight focus could lead to experimental challenges in resolving the vortices with a finite pixel-based detector. The dependence of the vortex radial position on the radial wave vector of the Bessel component illustrates that the mutual interaction between the two vortices is not primarily driven by the separation distance but rather more fundamentally by $k_{\rho}$ or $\theta$. For a given $k_{\rho}$, one can increase the number of braids by increasing the beam waist, because a larger $z_{max}$ yields a longer-lived composite Bessel-Gaussian. The numerical aperture dependence was confirmed by placing an iris at the imaging plane in Figure \ref{fig:schematic}(a) and observing the truncated $z_{max}$ and hence a smaller number of braids. For the idealized case of Figure \ref{fig:limitation}(c), the number of braids is infinite.

In summary, we have theoretically proposed optical vortex braiding by superposing Bessel modes with a plane wave, and we have experimentally demonstrated this braided structure by summing a Bessel-Gaussian beam and a coaxial Gaussian. Analytical predictions for braiding trajectory and period are consistent with experimental measurements. To the best of our knowledge, this is the first experimental demonstration of optical vortices rotating more than one revolution about a central axis.

\section*{Funding}

W. M. Keck Foundation and National Science Foundation (NSF, 1553905).

\section*{Disclosures}

The authors declare no conflicts of interest.

\bibliography{Refs}

\end{document}